\documentclass[preprint,superscriptaddress,preprintnumbers,showpacs]{revtex4}
\usepackage{amssymb}
\usepackage{graphicx}
\usepackage{dcolumn}
\usepackage{bm}
\usepackage{slashed}

\begin{document}

\newcommand*{\sjtu}{INPAC, SKLPPC and Department of Physics, Shanghai Jiao Tong University,  Shanghai, China}\affiliation{\sjtu}
\newcommand*{\NTU}{CTS, CASTS and Department of Physics, \\National Taiwan University, Taipei, Taiwan}\affiliation{\NTU}
\newcommand*{\ncts}{Department of Physics, National Tsing Hua University, and National Center for Theoretical Sciences, Hsinchu, Taiwan}\affiliation{\ncts}

\hfill{}

\title{Further Studies of Higgs Properties at An ILC  $\gamma\gamma$ Collider}
\affiliation{\NTU}
\author{Xiao-Gang He}\email{hexg@phys.ntu.edu.tw}\affiliation{\sjtu}\affiliation{\NTU}\affiliation{\ncts}
\author{Siao-Fong Li}\affiliation{\NTU}
\author{Hsiu-Hsien Lin}\affiliation{\NTU}

\begin{abstract}
Recently the ATLAS and CMS experiments at the LHC have found a Higgs like boson $h$ with a mass around 125 GeV from several decay modes.  The decay mode  $h \to \gamma\gamma$ is one of the most important modes in studying whether $h$ is actually the standard model (SM) Higgs boson. Current data indicate that $h\to \gamma\gamma$ has a branching ratio larger than the SM prediction for $h$ being identified as the SM Higgs boson. To decide whether the $h$ discovered at the LHC is the SM Higgs boson, more data are needed. We study how $\gamma\gamma$ collider can help to provide some of the most important information about the Higgs boson properties. We show that a $\gamma\gamma$ collider can easily verify whether the enhanced $h \to \gamma\gamma$ observed at the LHC hold. Different models can be tested by studying Higgs boson decay to $\gamma Z$.  Studying angular distribution of the $\gamma \gamma$ through on-shell production of $h$ and its subsequent decays into a $\gamma \gamma$ pair can decide whether the Higgs like boson $h$ at the LHC is a spin-0 or a spin-2 boson.

\end{abstract}


\date{\today $\vphantom{\bigg|_{\bigg|}^|}$}

\maketitle

The two detectors ATLAS~\cite{ATLAS} and CMS~\cite{CMS} at the LHC have discovered a boson particle $h$ with a mass about 125 GeV. By studying several decay modes~\cite{ATLAS,CMS,Chatrchyan:2012jja, atlas-note}, it has been shown that this particle matches many of the properties of the Higgs boson in the standard model (SM) from Higgs mechanism~\cite{A. Salam., Djouadi}. However, both ATLAS and CMS have found an enhanced $h \to \gamma \gamma$ branching ratio with a factor $R_{\gamma\gamma} = 1.8\pm0.5(ATLAS)$~\cite{ATLAS} ($1.6\pm0.4(CMS)$~\cite{CMS}) compared with that of the the SM prediction~\cite{Djouadi}. This prompted many theoretical studies of Higgs boson beyond the SM.
More data is required to confirm whether the enhancement is true or just an experimental fluctuation. The LHC will continue to run and obtain more data. It is expected that data from near future will be able to decide whether the $h$ boson discovered at the LHC is indeed the SM Higgs boson.  It is interesting to see if Higgs boson properties can be further studied with high precisions at different facilities to confirm the LHC results. To this end we note that a $\gamma\gamma$ collider is an idea place to study some of the most important properties of a Higgs boson. It can easily verify whether the enhanced $h \to \gamma\gamma$ observed at the LHC hold. Different models can be tested by studying Higgs boson decay to $\gamma Z$. Studying angular distribution of the final $\gamma \gamma$ through on-shell production of $h$ and its subsequent decays into a $\gamma \gamma$ pair can decide whether the Higgs like boson $h$ at the LHC has a spin-0 or spin-2.
\\

\noindent
{ \bf $\gamma\gamma \to h \to \gamma\gamma, \;\;\gamma Z$}
\\

Assuming that the $h$ boson discovered at the LHC is a spin-0 scalar particle, the matrix element $M(J=0, \gamma\gamma)$ of a spin-0 scalar coupling to $\gamma\gamma$ has the form
\begin{eqnarray}
M(0,\gamma\gamma) ={A}(k_{2\mu}k_{1\nu}-g_{\mu\nu}k_{1}\cdot k_{2}){\varepsilon^{\mu*}(k_{1})}{\varepsilon^{\nu*}(k_{2})}.
\end{eqnarray}
The decay width for $h \to \gamma\gamma$ is given by, $\Gamma_{0,\gamma\gamma} ={{A^{2}}m^3_h/{64\pi}}$.

If $h$ is the SM Higgs boson~\cite{Djouadi},
\begin{eqnarray}
&&A={{ie^2}\over{8\pi^2}}(\sqrt{2}G_{\mu})^{1\over2}[\Sigma_{f}N_{c}Q^{2}_{f}A_{1/2}^{H}(\tau_{f})+A_{1}^{H}(\tau_{W})]\;,\nonumber\\
&&A^{H}_{1/2}(x)={2[x+(x-1)f(x)]x^{-2}}\;,\\
&&A^{H}_{1}(x)={-[2x^{2}+3x+3(2x-1)f(x)]x^{-2}}  \;,\nonumber
\end{eqnarray}
where $N_c$ is the number of color, $\tau_i = {{m_{h}^{2}}/{4m^{2}_{i}}}$.
The definition of loop function $f(x)$ is
\[ f(x) = \left\{ \begin{array}{ll}
         {\arcsin^{2}\sqrt{x}} & \mbox{if $x\leq 1 $};\\
         {{-1\over4}[\ln({{1+\sqrt{1-x^{-1}}}\over{1-\sqrt{1-x^{-1}}}})-i\pi]^{2}}& \mbox{if $x > 1$}.\end{array} \right. \]

The cross section $\sigma(s)_{0,\gamma\gamma}$ for producing an on-shell $h$ at a monochromatic $\gamma\gamma$ collider followed by $h$ decays into a $\gamma\gamma$ pair with a center of mass (CM) frame energy $\sqrt{s}$ is directly related to the decay rate $\Gamma_{0, \gamma \gamma}$~\cite{Gunion:1992ce}. We have
\begin{eqnarray}
\sigma(s)_{0, \gamma\gamma} = \Gamma^{2}_{0, \gamma\gamma}{{8\pi^{2}}\over{m_{h}\Gamma_{total}}}\delta(s-m^2_h)\;,
\end{eqnarray}
where $\Gamma_{total}$ is the total $h$ decay width.

Realistically a $\gamma\gamma$ collider can be constructed by  using the laser backscattering
technique on the electron and positron beams in an $e^+e^-$ collider. For example the $e^+e^-$ ILC collider. Such a collider has been shown to be useful to study beyond SM physics~\cite{He:2006yy}. In this case the energy $E_\gamma$ of the photons are not monochromatic, but have a distribution $F(x)$ for a given electron/positron energy $E_e$~\cite{I. Ginzburg}
\begin{eqnarray}
&&F(x) = {1\over D(\xi)}(1-x + {1\over 1-x} - {4x\over \xi(1-x)} + {4x^2\over \xi^2(1-x)^2})\;,\nonumber\\
&&D(\xi) = (1-{4\over \xi} - {8\over \xi^2})\ln(1+\xi) + {1\over 2} + {8\over \xi} - {1\over 2 (1+\xi)^2}\;.
\end{eqnarray}
Here $x = E_\gamma/E_e$, and $\xi = 2(1+\sqrt{2})$.

To probe $h \to \gamma\gamma$ coupling at the $\gamma\gamma$ collider, one can study
$\gamma\gamma \to h \to \gamma\gamma$.  This cross section is dominated by on-shell $h$ production (narrow width approximation).
Convoluting the energies of the two photon beams produced by using the laser backscattering
technique on the electron and positron beams in an $e^+e^-$ collider with CM frame energy $\sqrt{s}$, we obtain for $\gamma\gamma\to h \to \gamma\gamma$ cross section $\sigma^L_{0, \gamma\gamma}(s)$
\begin{eqnarray}
\sigma(s)_{0, \gamma\gamma}^L = \int^{x_{max}}_{x_{min}} d x_1 \int^{x_{max}}_{x_{min}}dx_2 \sigma(x_1x_2 s)_{0, \gamma\gamma}F(x_1)F(x_2) = I(m^2_h/s){{8\pi^{2}}\over{m^{3}_{h}}}\Gamma_{0, \gamma\gamma}B_{0, \gamma\gamma}\;,\label{cross}
\end{eqnarray}
where, $B_{0, \gamma\gamma} = \Gamma_{0, \gamma\gamma}/\Gamma_{0, total}$. In the SM $B_{\gamma\gamma} =2.28\times 10^{-3}$~\cite{Dittmaier:2012vm}. The function $I(y)$ is~\cite{He:2006yy}
\begin{eqnarray}
I(y) = \int^{x_{max}}_{y/x_{max}} d x {y\over x} F(x)F(y/x)\;,
\end{eqnarray}
with $y = m^2_h/s$,  $x_{max} = \xi/(1+\xi)$, and $x_{min} = y/x_{max}$.
Note that the function $I(y)$ is a function of $m^2/s$ only. The model-dependent part resides
purely in the expression for $\Gamma_{0, \gamma\gamma}B_{0, \gamma\gamma}$. In Fig. 1 we show $I(y)$ as a function of y. We see
that for a large range of $m^2_h/s$, $I(y)$ is sizeable for $m_h = 125$ GeV with $\sqrt{s}$ in the range 160  to 320 GeV.
A $e^+e^-$ collider with a $\sqrt{s}$ in this range can be very useful for the purpose of studying $h$ properties. When energy becomes higher the cross section goes down.

It is clear from eq. (\ref{cross}) that $\gamma\gamma$ collider can easily provide information about $h \to \gamma\gamma$ coupling. In Fig. 2, we show the SM for $\sigma(s)_{0, \gamma\gamma}^L$ as a function of $\sqrt{s}$. Deviation from this prediction is an indication of new physics beyond the SM. In the 160 to 320 GeV range for $\sqrt{s}$, with an integrated luminosity of 500 fb$^{-1}$ for an ILC, one would have 45  to 70 number of events. This can provide important information.  The angular distribution may also be studied with such a number of events.

If the current LHC data for enhanced $h \to \gamma\gamma$ hold, one would obtain a $\sigma_{0, \gamma\gamma}^L(s)$ which is larger than SM prediction by a factor $R_{\gamma\gamma}^2$ if the total decay width is the same as that of SM prediction. If the branching ratio is kept the same as that of SM, then the enhancement for $\sigma(s)^L_{0, \gamma\gamma}$ is $R_{0, \gamma\gamma}$. If $\Gamma_{0, \gamma\gamma} B_{0, \gamma\gamma}$ turns out to be the same as that of SM prediction, but  $\Gamma_{0, \gamma\gamma}$ and  $B_{0, \gamma\gamma}$ are different from those of SM predictions, information from other modes is needed to see whether there is new physics beyond the SM.

To see how information from additional decay modes can help to distinguish different models, let us take $h \to \gamma Z$ for example.
In this case, one obtains the cross section $\sigma(s)_{0, \gamma Z}^L$ for $\gamma\gamma \to h\to \gamma Z$ as
\begin{eqnarray}
\sigma(s)_{0, \gamma Z}^L  = I(m^2_h/s){{8\pi^{2}}\over{m^{3}_{h}}}\Gamma_{0, \gamma \gamma} B_{0, \gamma Z}\;,
\end{eqnarray}
where $B_{0, \gamma Z}$ is the branching ratio for $h \to \gamma Z$.

Taking the ratio of $\sigma(s)_{0, \gamma\gamma}^L$ and $\sigma(s)_{0, \gamma Z}^L$, we have
\begin{eqnarray}
S_{0, \gamma Z/\gamma\gamma} = {\sigma(s)_{0, \gamma Z}^L\over \sigma(s)_{0, \gamma \gamma}^L} = {\Gamma_{0, \gamma Z}\over \Gamma_{0, \gamma\gamma}}\;.
\end{eqnarray}

In the SM, $S_{0, \gamma Z/\gamma\gamma}$ is predicted to be 0.71~\cite{Dittmaier:2012vm}. Again if the LHC data for enhancement of $ h\to \gamma\gamma$ hold,
new physics is need to produce the enhancement factor which may also modify $\Gamma_{0, \gamma Z}$.
The measurement of $S_{0, \gamma Z/\gamma\gamma}$ can provide important information about Higgs properties in the SM and also beyond. $h \to \gamma Z$ has not been measured at the LHC. When it is measured, the expected value for $S_{0, \gamma Z/\gamma\gamma}$ to be measured at a $\gamma\gamma$ collider is determined. The $\gamma\gamma$ collider can provide additional confirmation.
\\

\begin{figure}
\includegraphics[width=10cm]{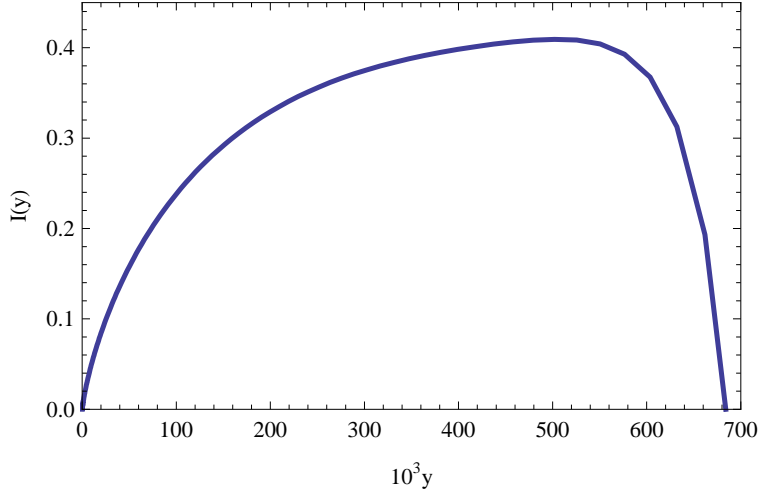}
\caption{$I(y)$ as a function of $y$.}\label{fig1}
\end{figure}

\begin{figure}
\includegraphics[width=10cm]{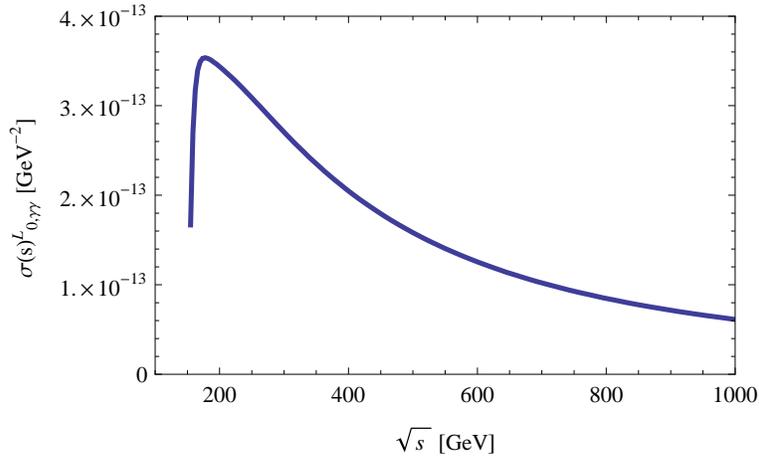}
\caption{$\sigma(s)^L_{0, \gamma\gamma}$ (in unit of GeV$^{-2}$) as a function of $\sqrt{s}$ in the SM.}\label{fig2}
\end{figure}

\noindent
{\bf Angular distribution of $\gamma$ and spin of $h$ at a $\gamma\gamma$ collider}
\\

\begin{figure}
\includegraphics[width=6cm]{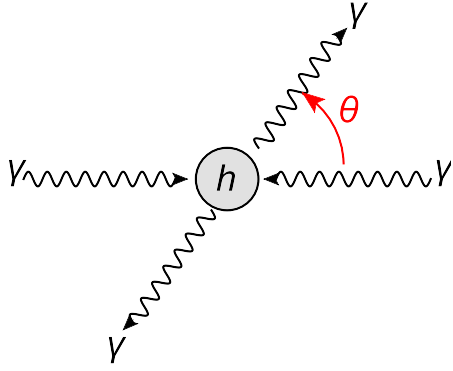}
\caption{The angle $\theta$ of a final photon in $\gamma\gamma \to \gamma\gamma$.}\label{fig3}
\end{figure}

Since the LHC has observed $h \to \gamma\gamma$, the $h$ boson cannot be a spin-1 particle due to Landau-Yang theorem~\cite{Landau}. To identify $h$ to be the SM Higgs boson, one needs to have precise information about the spin of the $h$ boson. Several papers have discussed implications of $h$ with a different spin than zero~\cite{Choi, spin2}.
There have been analyzed based on LHC data on the $h$ boson spin favoring the spin to be
zero~\cite{atlas-note}.
One of the interesting ways to obtain information about $h$ spin at the LHC is to use  $gg\rightarrow h\rightarrow\gamma\gamma$ angular distribution which has been considered in Ref.~\cite{Choi}. At the LHC there are also
contributions from $q\bar{q} \to h \to \gamma\gamma$ process contaminating the analysis. With a $\gamma\gamma$ collider one can provide more controlled information about $h$ spin because the cross section for $e^+e^- \to h \to \gamma\gamma$ is extremely small.

For a scalar $h$, the angular distribution for one of the final photon respecting to the incoming $\gamma$ beams shown in Fig. 3 is isotropic in the $\gamma\gamma$ CM frame~\cite{Choi}
\begin{eqnarray}
{1\over \sigma(s)_{0, \gamma\gamma}}{d\sigma(s)_{0, \gamma\gamma}\over d \cos\theta} = 1\;. \label{dis}
\end{eqnarray}

In the $e^+e^-$ CM frame (laboratory frame), collision of the two photons  is not in the $\gamma\gamma$ CM frame and therefore
the distribution of the photons is not the same as that predicted by eq. (\ref{dis}). In the laboratory frame, depending on the values of $x_{1}$ and $x_{2}$, the two photons may have different energy. The $h$ produced will be boosted to the direction of the photon with a larger $x_i$. The angle $\theta$ when seeing from laboratory frame will be changed to $\theta_L$. The relation between $\theta$ and $\theta_L$ can be easily shown to be
\begin{eqnarray}
\cos \theta = \frac{\cos \theta_{L}+\beta}{\beta\cos \theta_{L}+1}\;,\;\;\frac{d\cos \theta}{d\cos \theta_{L}} = \frac{1-\beta ^{2}}{(\beta\cos \theta_{L}+1)^{2}}\;,\;\;\beta = \frac{x_{1}-x_{2}}{x_{1}+x_{2}}\;.
\end{eqnarray}

The laboratory frame angular distribution $A(0,\theta_L)$ of $\theta_{L}$ for a spin-0 scalar can be studied by the following convoluted distribution,
\begin{eqnarray}
A(0, \theta_L) = {1\over \sigma(s)_{0, \gamma\gamma}^L} \int^{x_{max}}_{x_{min}} d x_1 \int^{x_{max}}_{x_{min}}dx_2  F(x_1)F(x_2) {d\sigma(s)_{0, \gamma\gamma}\over d \cos\theta_{L}}\;.
\end{eqnarray}

One can also defined a similar quantity for the case of the $h$ boson being a particle with a different spin, spin-J, $A(J, \theta_L)$. We find that this quantity can give information about the spin of the $h$ boson.
To see how this works, we take an example of a spin-2  $h$ tensor coupled to $\gamma\gamma$ to study $A(2,\theta_L)$ and compared with $A(0,\theta_L)$ for a spin-0 scalar.

The matrix element of a spin-2 tensor coupling to $\gamma\gamma$ can be written as~\cite{Han:1998sg}
\begin{eqnarray}
&&M(2, \gamma\gamma) = \frac{-\kappa}{2}[(k_{1}\cdot k_{2})C_{\mu\nu,\varrho\sigma}+D_{\mu\nu,\varrho\sigma}(k_{1},k_{2})]\varepsilon ^{\rho *}(k_{1})\varepsilon ^{\sigma *}(k_{2})\epsilon^{\mu\nu }\;,\nonumber\\
&&C_{\mu\nu,\varrho\sigma}=\eta_{\mu\varrho}\eta_{\nu\sigma}+\eta_{\mu\sigma}\eta_{\nu\rho}
-\eta_{\mu\nu}\eta_{\rho\sigma}\;,\\
&&D_{\mu\nu,\varrho\sigma}(k_{1},k_{2}) = \eta_{\mu\nu}k_{1\sigma}k_{2\rho}-[\eta_{\mu\sigma}
k_{1\nu}k_{2\rho}+\eta_{\mu\rho}k_{1\sigma}k_{2\nu}-\eta_{\rho\sigma}k_{1\mu}k_{2\nu}+
(\mu\leftrightarrow\nu)]\;.\nonumber
\end{eqnarray}

The matrix element for $\gamma \gamma \to h \to \gamma\gamma$ is given by
\begin{eqnarray}
&&M(2, \gamma\gamma \to h \to\gamma\gamma) =\frac{-\kappa^{2}}{8} \varepsilon ^{\rho }(k_{1})\varepsilon ^{\sigma }(k_{2})\varepsilon ^{\gamma*}(q_{1})\varepsilon ^{\delta*}(q_{2})\nonumber\\
&&\times \frac{[((k_{1}\cdot k_{2})C_{\mu\nu,\varrho\sigma}+D_{\mu\nu,\varrho\sigma}(k_{1},k_{2}))B^{\mu\nu,\alpha\beta}((q_{1}\cdot q_{2})C_{\alpha\beta,\gamma\delta}+D_{\alpha\beta,\gamma\delta}(q_{1},q_{2}))]}{s-m_{h}^{2}}\;,
\end{eqnarray}
where $B_{\mu\nu,\alpha\beta}(k) = (\eta_{\mu\alpha}-\frac{k_{\mu}k_{\alpha}}{m_{h}^{2}})(\eta_{\nu\beta}-\frac{k_{\nu}k_{\beta}}{m_{h}^{2}})+(\eta_{\mu\beta}-\frac{k_{\mu}k_
{\beta}}{m_{h}^{2}})(\eta_{\nu\alpha}-\frac{k_{\nu}k_{\alpha}}{m_{h}^{2}})-\frac{2}{3}(\eta_{\mu\nu}-\frac{k_{\mu}k_{\nu}}{m_{h}^{2}})
(\eta_{\alpha\beta}-\frac{k_{\alpha}k_{\beta}}{m_{h}^{2}})$.

This gives a differential cross section $d\sigma(s)_{2,\gamma\gamma}/d\cos\theta$ for $\gamma \gamma \to h \to \gamma\gamma$, in the narrow width approximation, as
\begin{eqnarray}
&&\frac{d\sigma(s)_{2,\gamma\gamma}}{d\cos\theta} = \frac{25\pi^{2}}{2m_{h}}(\cos ^ {4} \theta +6 \cos ^ {2} \theta + 1)\Gamma_{2, \gamma\gamma}B_{2, \gamma\gamma}\delta(s-m_{h}^{2})\;,\nonumber\\
&&\Gamma_{2, \gamma\gamma} =\frac{\kappa^{2} m_{h}^{3}}{320\pi}\;.
\end{eqnarray}
In the above to obtain the expression for $\Gamma_{2, \gamma\gamma}$, we have summed the polarization $\lambda$ of the spin-2 tensor $\epsilon^\lambda_{\mu\nu}$ with
$\sum_\lambda \epsilon^{\lambda *}_{\mu\nu}\epsilon^\lambda_{\rho\sigma} = ({1}/{2})B_{\mu\nu,\rho\sigma}(k)$~\cite{Han:1998sg}.

Integrating the angle $\theta$ out, we have
\begin{eqnarray}
\sigma(s)_{2,\gamma\gamma}&=&\frac{8\pi^{2}(2J+1)}{m_{h}}\Gamma_{J, \gamma\gamma}B_{J, \gamma\gamma}\delta(s-m_{h}^{2})\;.
\end{eqnarray}
In our case $J = 2$. The above equation works for any spin $J$.

In the $\gamma\gamma$ CM frame, we have
\begin{eqnarray}
{1\over \sigma(s)_{2,\gamma\gamma}}{d\sigma(s)_{2,\gamma\gamma}\over d\cos\theta} ={5\over 16}(\cos^4\theta+ 6\cos^2 \theta + 1)\;.
\end{eqnarray}

Note that even in the $\gamma\gamma$ CM frame, the final $\gamma$ has a non-trivial angular distribution for spin-2 tensor. This fact can be used to distinguish whether $h$ is a spin-0 or a spin-2 particle. One might wonder that the expression for $\sigma(s)_{i,\gamma\gamma}$ already provides enough information to distinguish whether $h$ is a spin-0 and spin-2 particle. This may not be sufficient if $\Gamma_{i, \gamma\gamma}$ and $B_{i, \gamma\gamma}$ are not separately measured. If separately measured,  $\sigma(s)_{2,\gamma\gamma}$ would be five times larger than $\sigma(s)_{0, \gamma\gamma}$.  We find that if comparing $A(0, \theta_L)$ and $A(2,\theta_L)$, one can distinguish different cases even without knowing  $\Gamma_{i, \gamma\gamma}$ and $B_{i, \gamma\gamma}$  separately.

In Fig. 4, we plot $A(0, \theta_L)$ and $A(2,\theta_L)$ for several different $\sqrt{s}$. We see that at
$\sqrt{s} = 200$ GeV, the difference for spin-0 and spin-2 laboratory frame angular distribution are substantial and can easily distinguish these two cases. In fact in the range of $160  \sim 320$ Gev for $\sqrt{s}$, where the function $I(y)$ is sizeable (high event number), the difference between spin-0 and spin-2 are all substantial and can provide information about spin. When energy increases, the boosting effect dominates over the intrinsic angular distribution, the difference becomes smaller making the distinction difficulty. To study the spin of $h$ boson of mass 125 GeV, the energy range $\sqrt{s}$ in the range 160 to 320 GeV is better than energies in higher ranges.

To summarize, we have studied how a $\gamma\gamma$ collider can help to provide some of the most important information about the Higgs boson properties which the LHC have/will found. We have shown that a $\gamma\gamma$ collider can easily verify whether the enhanced $h \to \gamma\gamma$ observed at the LHC hold. Different models can be tested by studying Higgs boson decay to $\gamma Z$.  Studying angular distribution of the $\gamma \gamma$ through on-shell production of $h$ and its subsequent decays into a $\gamma \gamma$ pair can decide whether the Higgs like boson $h$ at the LHC has a spin-0 or spin-2.

\begin{figure}
\includegraphics[width=8cm]{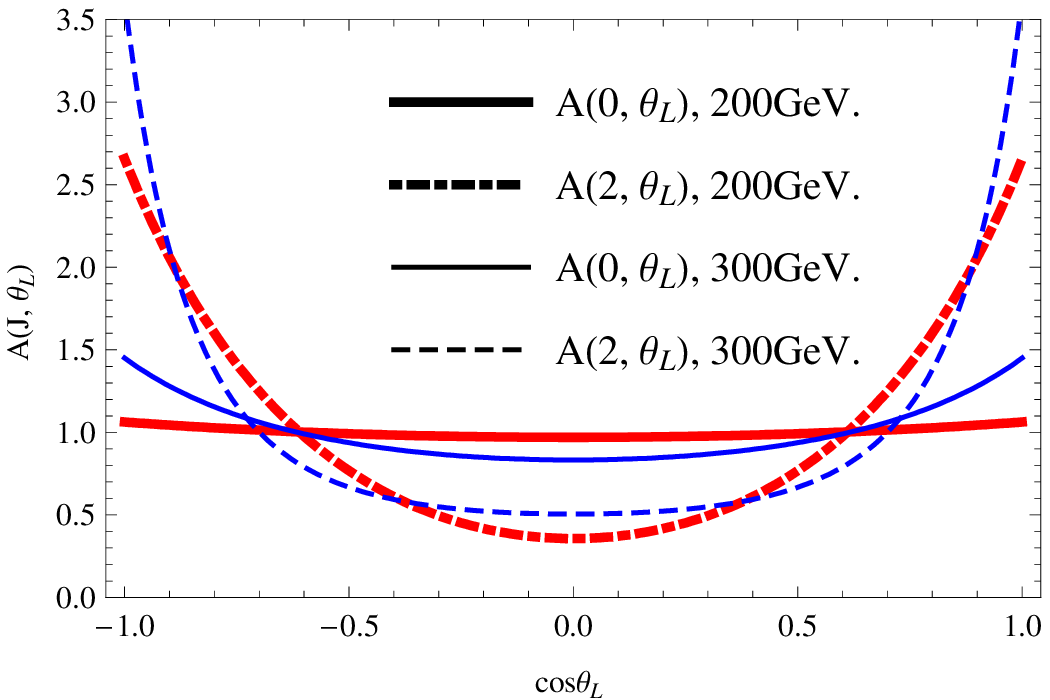}
\includegraphics[width=8cm]{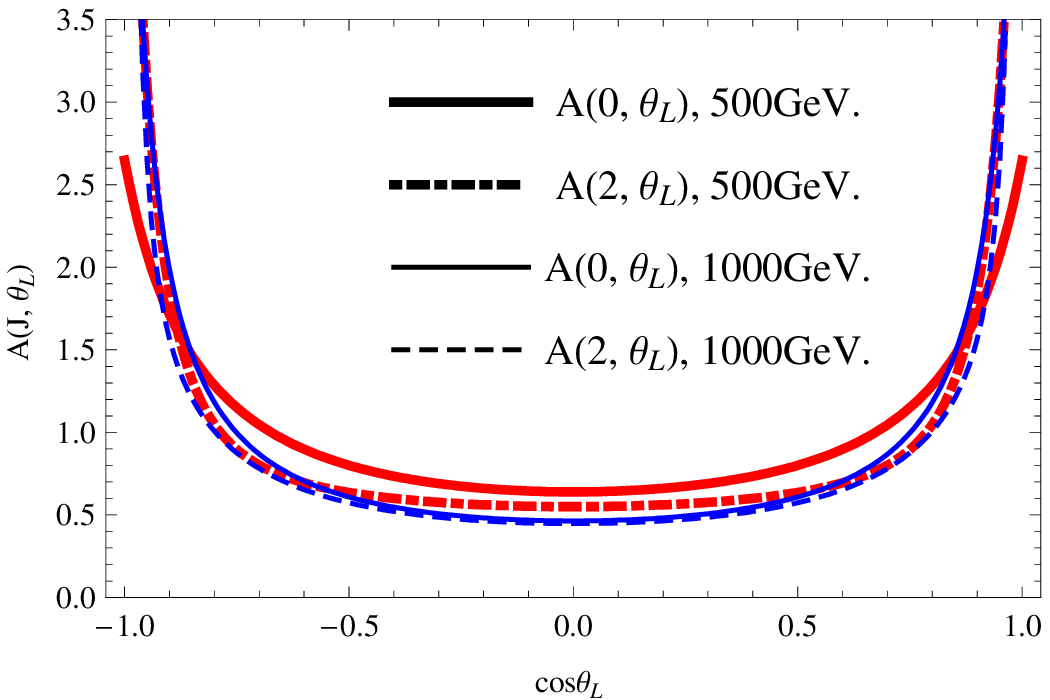}
\caption{Angular distribution $A(J=0,2, \gamma\gamma)$ in $\gamma\gamma \to h \to \gamma \gamma$ with different energies $\sqrt{s}$.}\label{fig3}
\end{figure}

\acknowledgments \vspace*{-1ex}
This work was supported in part by NSC of ROC, and NNSF(grant No:11175115) and Shanghai science and technology commission (grant No: 11DZ2260700) of PRC.

\end{document}